\documentclass[aps,prl,twocolumn,superscriptaddress]{revtex4}
\usepackage{mathrsfs}
\usepackage{epsfig}
\usepackage{graphicx}
\usepackage{amsfonts}
\usepackage[figuresright]{rotating}
\usepackage{amssymb}
\usepackage{amsmath}
\usepackage{dcolumn}
\usepackage{bm}
\usepackage{color}

\def\be{\begin{equation}} \def\ee{\end{equation}}
\def\bea{\begin{eqnarray}} \def\eea{\end{eqnarray}}

\newcommand{\WQCASQC} { Wilczek Quantum Center and Key Laboratory of Artificial Structures and Quantum Control, School of Physics and Astronomy, Shanghai Jiao Tong University, Shanghai 200240, China}

\newcommand{\tdli}{T. D. Lee Institute, Shanghai Jiao Tong University and Shanghai Research Center for Quantum Sciences, Shanghai 200240, China}

\begin{document}
\title{Dimension-raising phase transitions in driven magnets and condensates}

\author{Zhizhen Chen}
\affiliation{\tdli}
\affiliation{\WQCASQC}

\author{Zi Cai}
\email{zcai@sjtu.edu.cn}
\affiliation{\WQCASQC}

\begin{abstract}  

We propose a periodically driven system whose dimensionality is an emergent property that can be tunable, thus enables us to realize not only many-body phases with arbitrary dimensions, but also phase transitions, instead of crossovers, between phases with various dimensions. We study an interacting rotor model whose instantaneous Hamiltonian keeps the one-dimensional (1D) feature at any given time. Despite this, an emergent two-dimensional (2D) phase appears when the driving frequency exceeds a critical value, at which a dimension-raising phase transition takes place. We find that the nonequilibrium feature of the system could qualitatively change the finite temperature critical behavior of the emergent 2D phase and make it different from its equilibrium counterpart.  A four-dimensional (4D) generalization and experimental realizations of the proposed model  based on a programmable reconfiguration technique in optical tweezers setups have also been discussed.

\end{abstract}


\maketitle

{\it Introduction --} Dimensionality is a key factor that influences the properties of many-body systems, since it profoundly modifies the effects of  fluctuations.  Although the dimensionality is usually considered to be a well-fixed quantity, an important class of systems exists in solid state and cold atom, where the dimensionality itself becomes a tunable and emergent property. These systems exhibit intriguing phenomena including a dimensional crossover with a drastic change of the properties tied to the dimensionality\cite{Klanjsek2008,Vogler2014,Revelle2016,Moller2021,Yao2023,Shah2023,Kirankumar2024}, or a dimensional reduction with low-dimensional physics spontaneously emerging from systems with higher dimensions\cite{Khomskii2005,Streltsov2017,Sebastian2006,Batista2007,Okuma2021,Feng2024}. Several important questions arises in these regards, for instance, is it possible to realize a genuine phase transition, instead of crossover, between the phases of matter with different dimensions?  The existence of dimensional reduction naturally raise the question whether the opposite way is also possible, that the effective dimension of a system is spontaneously increased instead of being reduced. Whether it is possible to realize many-body phases with effective dimensions higher than three?

To address these issues, we propose a dimension-raising mechanism via a periodic driving protocol. In the past decades, time-periodic fields have been used as a knob to manipulate many-body phases and phase transitions in eletronic materials\cite{Oka2009,Lindner2011,Zhou2023}, cold atoms\cite{Struck2011,Jotzu2014} and  magnetic systems\cite{Tome1990,Sides1998,Wan2017}. Recent developments in synthetic quantum materials enable us to dynamically modulate key ingredients of many-body systems other than external fields, such as  interaction\cite{Hu2019,Feng2019,Zhang2020,Yue2023,Fu2024}, lattice geometry and connectivity\cite{Ebadi2022,Bluvstein2023}. The latter plays a remarkable role in neutral atom quantum computation, {\it e.g.} the Rydberg atoms are trapped by array of optical tweezers that can be individually moved during the computation while preserving the quantum coherence\cite{Bluvstein2023}. Such a reconfigurability of dynamically tuning the lattice geometry and connectivity has provided new opportunities for the studies of nonequilibrium many-body physics.

\begin{figure}[htbp]
\centering
\includegraphics[width=0.46\textwidth]{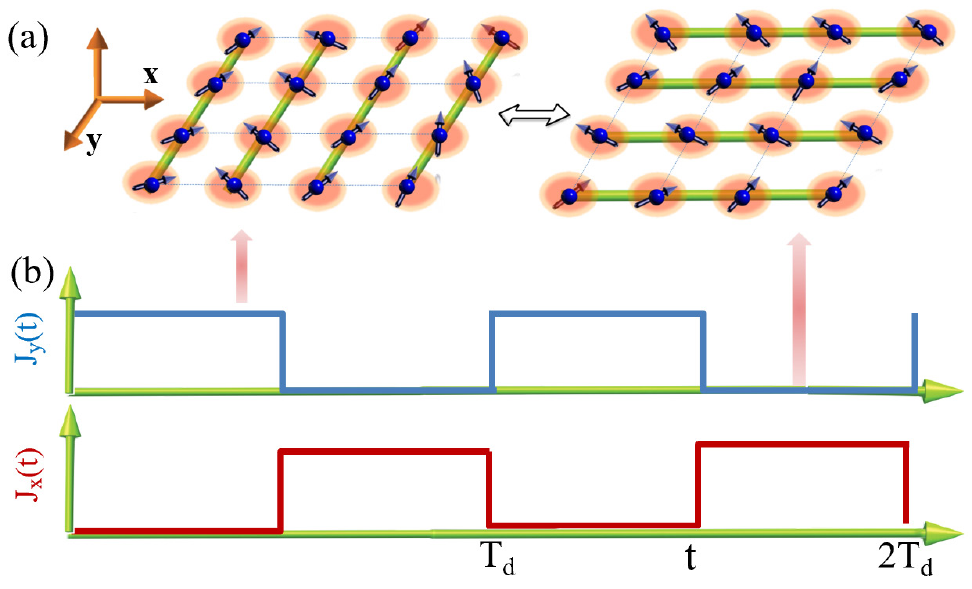}
\caption{(a) Sketch of the interacting rotor model whose couplings are periodically switched between the patterns with either horizontal bonds or vertical bonds only.   (b) The periodic driving protocol for $J_x(t)$ and $J_y(t)$.}
\label{fig:fig1}
\end{figure}

\begin{figure*}[htb]
\includegraphics[width=0.99\textwidth]{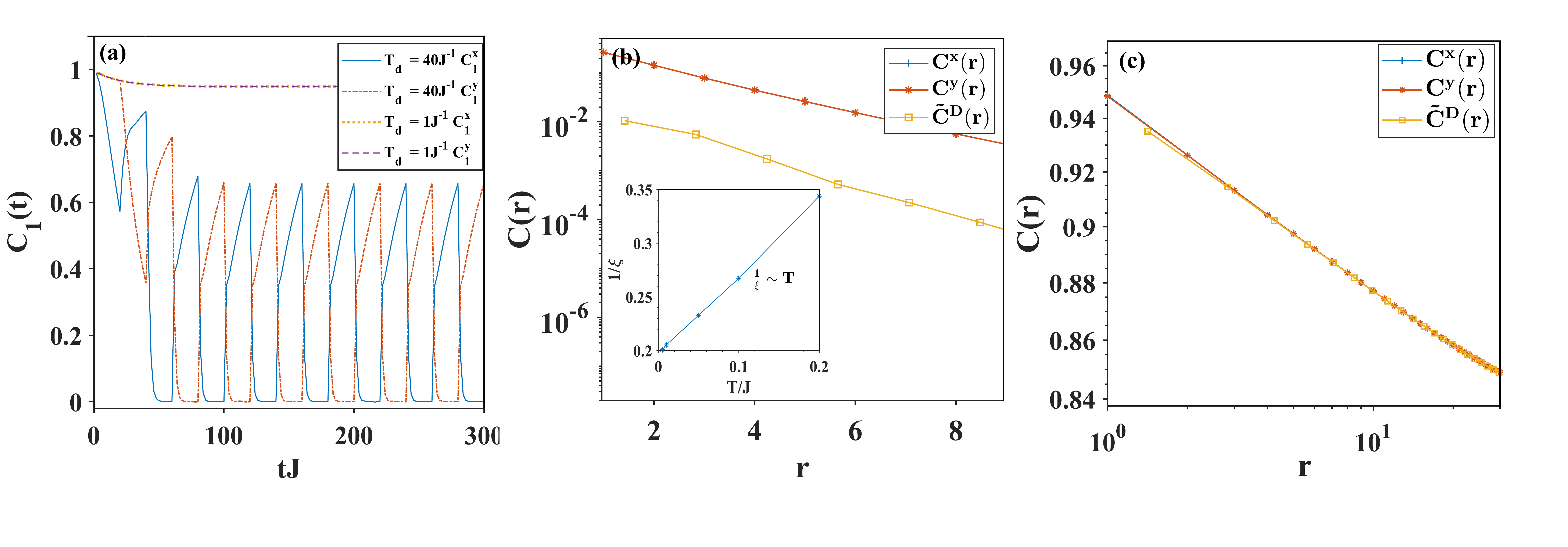}
\caption{(a) The dynamics of the NN correlation along the x and y directions ($C_1^x(t)$ and $C_1^y(t)$) in the presence of slow ($T_d=40J^{-1})$ and fast ($T_d=J^{-1}$) driving respectively;  The time-averaged correlation function $C(\mathbf{r})$ in the presence of (b) slow driving  and (c) fast driving along the x, y and diagonal directions ($C^x(r)$, $C^y(r)$ and $\tilde{C}^D(r)=C^D(\sqrt{2}r)$, here the distance along the diagonal direction is normalized by a factor of $\sqrt{2}$ to make it the same as the distance along the x and y directions). The insets of (b) indicate the temperature dependence of the inverse of correlation length $1/\xi$ in the slowly driven case with $T_d=80J^{-1}$.   Other parameters are chosen as $T=0.1J$, $I=J$, $\gamma=0.05J$ and $N=100^2$.}
\label{fig:fig2}
\end{figure*}

In this study, we investigate an interacting rotor model in a lattice whose connectivity is periodically switched between the patterns with either horizontal bonds or vertical bonds only (Fig.\ref{fig:fig1} a), thus its instantaneous Hamiltonian at any given time is a set of decoupled 1D chains.  Despite its 1D feature, this system can exhibit an emergent dimension-raising phenomena that are uniquely tied to its nonequilibrium nature. By tuning the driving frequency, we find a nonequilibrium phase transition (NEPT)  from an asymmetric disordered phase, to an emergent isotropic 2D critical phase.  Although the latter resembles the quasi-long-range ordered phase of an equilibrium 2D XY model, we find that the periodic driving could qualitatively change the nature of its finite temperature phase transition from the Berezinski-Kosterlitz-Thouless(BKT) type\cite{Berezinskii1971,Kosterlitz1973} to the 1st order.   A higher-dimensional generalization of this driving protocol make it possible to realize many-body phases with arbitrary dimensions.

{\it Model and method --} The system we studied is composed of a set of interacting  rotors, and the instantaneous Hamiltonian of the system reads:
\begin{equation}
H(t)=\sum_\mathbf{i} \big\{\frac{I}2 \dot{\theta}^2_\mathbf{i} -\sum_{\alpha=x,y} J_\alpha (t)\cos(\theta_{\mathbf{i}}-\theta_{\mathbf{i}+\hat{e}_\alpha}) \big\} \label{eq:energy}
\end{equation}
where $\hat{e}_{x(y)}$ indicates the unit vector along the x(y) direction.  $\theta_\mathbf{i}$ is the polar angle of the $\mathbf{i}$th rotor, whose kinetic energy reads $\frac{I}2 \dot{\theta}^2_\mathbf{i}$ with $I$ being its rotational inertia. Each rotor interacts with its neighbors along the horizontal(vertical) direction via a time-dependent coupling strength  $J_{x(y)}(t)$, which takes the form:
\begin{small}
\begin{equation}
\begin{cases}
J_x(t)=J,~ J_y(t)=0;  &  nT_d <t\leq (n+\frac 12) T_d\\
J_x(t)=0,~ J_x(t)=J;  & ( n+\frac 12) T_d <t\leq (n+1) T_d
	\end{cases} \label{eq:Jxy}
\end{equation}
\end{small}
with n being an integer. $T_d$ is the period of the driving. Eq.(\ref{eq:Jxy}) represents a periodic driving protocol that at the 1st (2nd) half of each period, only the vertical (horizontal) coupling are switched on (see Fig.\ref{fig:fig1} b), thus the system Hamiltonian is always a set of decoupled rotor chains along either x or y direction during the evolution.

To balance the energy pumped by the driving, we assume each rotor is additionally coupled to a thermal bath, whose effect is modeled by the Langevin equation\cite{Wan2018}:
\begin{equation}
I\ddot{\theta}_\mathbf{i}+\gamma\dot{\theta}_\mathbf{i}=f_\mathbf{i}(t)+\xi_\mathbf{i}(t) \label{eq:EOM}
\end{equation}
where the mechanical torque $f_i(t)$ takes the form:
\begin{equation}
f_\mathbf{i}(t)=\sum_{\alpha=x,y} J_\alpha(t) \sin(\theta_\mathbf{i}-\theta_{\mathbf{i}+\vec{e}_\alpha}),
\end{equation}
 $\gamma$ indicates the strength of the bath-induced friction. The thermal noise $\xi_\mathbf{i}(t)$ is a zero-mean stochastic torque, which can be modeled by a white  noise: $\langle \xi_\mathbf{i}(t)\xi_\mathbf{j}(t')\rangle_\xi= \mathcal{D}^2 \delta_{\mathbf{ij}}\delta(t-t')$,
where $\langle\cdot\rangle_\xi$ indicates the ensemble average over different noise trajectories.  $\mathcal{D}$ is the strength of the noise and satisfies the classical fluctuation-dissipation theorem: $\mathcal{D}^2=2k_B T \gamma I$
where T is the temperature of the bath. Although the temperature of a non-equilibrium system is ill-defined, we neglect the back action of the system to the bath, which thus remains in thermal equilibrium with a fixed temperature during evolution.

Numerically, we solve the stochastic differential Eq.(\ref{eq:EOM}) using the  Heun  algorithm\cite{Ament2016} with a discrete time step of $\Delta t=10^{-2}J^{-1}$. We consider a lattice with  periodic boundary condition, and the simulated system size is up to $N=100^2$.  The ensemble average over the noise trajectories can be performed by directly sampling over a set of independent noise realizations. We will focus on the long-time asymptotic behaviors of the model and their dependence on the driving period $T_d$ and temperature $T$. To this end, throughout this paper we set $J$ to be unit and fix $\gamma=0.05J$ and $I=J$.

\begin{figure}[htbp]
\centering
\includegraphics[width=0.49\textwidth]{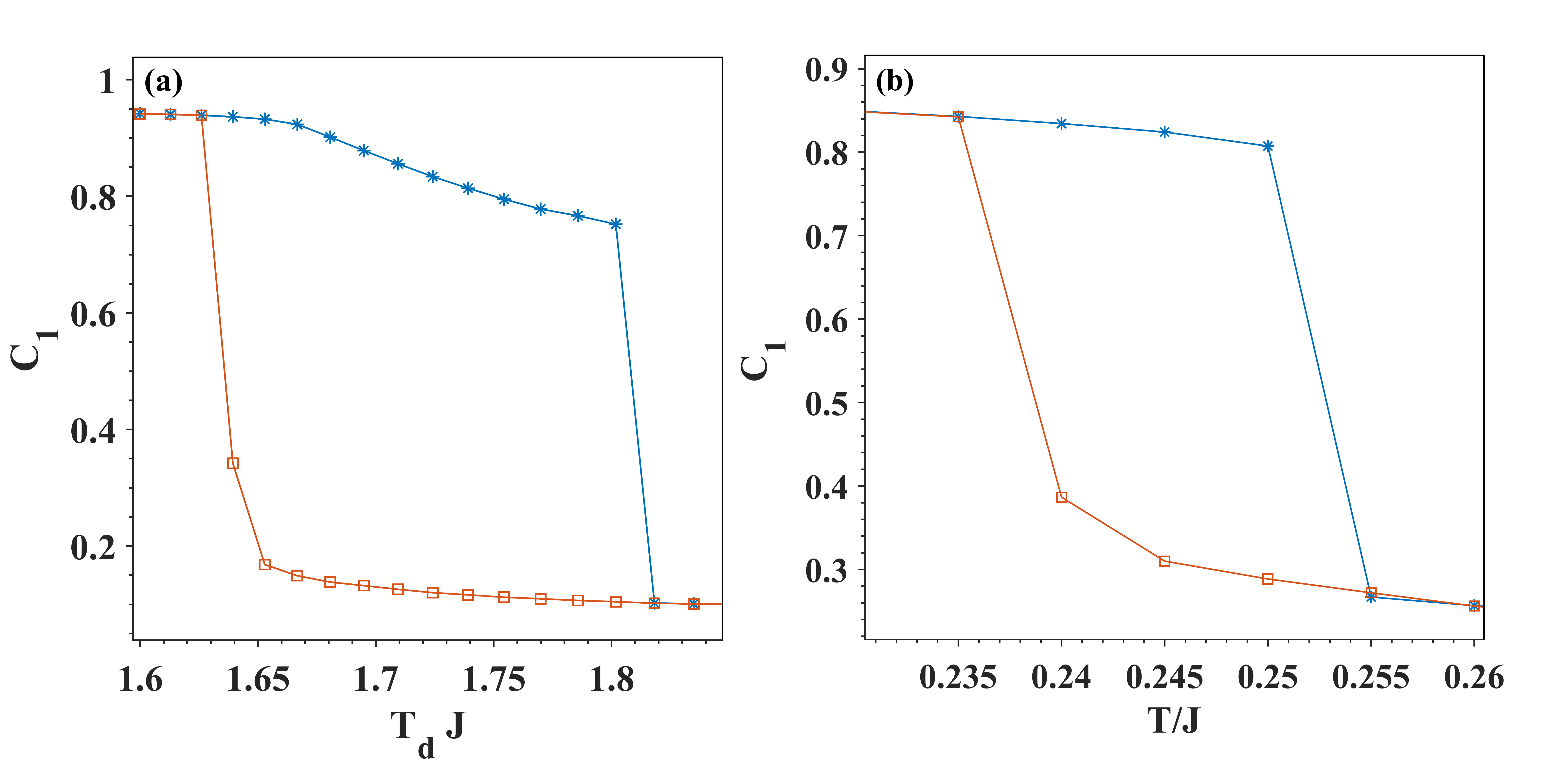}
\caption{The hysteresis loops in the (a) $C_1-T_d$ curves with a fixed $T=0.1J$ and (b) $C_1-T$ curves with fixed $T_d=1.54J^{-1}$. The red square (blue star) dots  indicate the results of simulations in the process of forward (backward) tuning of the control parameters. Parameters $I$, $\gamma$ and $N$ are chosen the same as in Fig.\ref{fig:fig2}.}
\label{fig:fig3}
\end{figure}

{\it Emergent dimension-raising isotropic phase --} To begin with, we focus on the dependence of the long-time asymptotic behavior on the driving period $T_d$ by fixing  $T=0.1J$. The driven-dissipative feature of our model give rise to two competing time scales: the driving period $T_d$ and the relaxation time $\tau$. In the adiabatic limit ($T_d\gg \tau$), the system adiabatically follows the thermal equilibrium states of the instantaneous 1D Hamiltonian.(\ref{eq:energy}). In the opposite limit of fast driving ($T_d\gg \tau$), the stroboscopic dynamics of this periodically driven system is governed by an time-independent effective Hamiltonian, which can be derived via the Magus expansion in the Floquet formalism. The zero-order of Magus expansion gives an  Floquet Hamiltonian $H_0=\frac{1}{T_d}\int_0^{T_d} dt H(t)$ as the time average over one period of Hamiltonian.(\ref{eq:energy}):
\begin{small}
\begin{equation}
H_0=\frac 12 \sum_\mathbf{i} \big\{I \dot{\theta}^2_\mathbf{i} -J[\cos(\theta_\mathbf{i}-\theta_{\mathbf{i}+\hat{e}_x})+\cos(\theta_\mathbf{i}-\theta_{\mathbf{i}+\hat{e}_y})] \big\} \label{eq:H0}
\end{equation}
\end{small}
which takes the same form of a 2D rotor model. Despite the simplicity of this analysis, the real situation is more complex. For example, the low temperature phase of $H_0$ is a critical phase whose relaxation time diverges ($\tau\rightarrow \infty$), which invalidates the condition of Magnus expansion ($T_d\gg \tau$) and calls for careful numerical analysis.

In a nonequilibrium system, the thermodynamic quantities such as the free energy, specific heat, etc are ill-defined, thus to characterize non-equilibrium phases and phase transitions, we calculate the equal-time correlation function $C(\mathbf{r},t)$ at time t, which is defined as
 \begin{equation}
 C(\mathbf{r},t)=\frac 1{L^2}\sum_\mathbf{i} \langle\cos[\theta_\mathbf{i}(t)-\theta_{\mathbf{i}+\mathbf{r}}(t)]\rangle_\xi
 \end{equation}
We start with the nearest neighboring (NN) correlation: $C_1^{x(y)}(t)=C(\mathbf{r},t)$ with $\mathbf{r}=\hat{e}_{x(y)}$.  The dynamics of $C_1(t)$ with slow and fast driving are plotted in Fig.\ref{fig:fig2} (a), which exhibit striking difference. At slow driving, the system exhibit a significant 1D feature characterized by the asymmetric feature along the x- and y-directions: at most of the time, either $C_1^x(t)$ or $C_1^y(t)$ vanishes.   In contrast, at fast driving, $C_1(t)$ barely oscillates, and the asymmetry between $C_1^x(t)$  and $C_1^y(t)$ almost vanishes.

 The phases and phase transitions are usually characterized by the long-range correlations. We define a time-averaged correlation as $C(\mathbf{r})=\frac 1{T'}\int_{t_0}^{t_0+T'} dt C(\mathbf{r},t)$, where $t_0$ is the relaxation time and $T'$ is the period over which the time average is performed. We define $C(\mathbf{r})$ along x, y and diagonal directions as $C^x(r)$, $C^y(r)$ and $C^D(r)$ with $\mathbf{r}=(r,0), (0,r)$ and $(r,r)$ respectively. Fig.\ref{fig:fig2} (b) indicates that at slow driving, all these correlation functions decay exponential over distance, and $C^D(r)$ is much smaller than $C^x(r)$ and $C^y(r)$, agreeing with the quasi-1D feature that the correlations along the directions other than x or y almost vanish.

 At fast driving, Fig.\ref{fig:fig2} (c) shows  $C^D(r)$ coincide with $C^x(r)$ and $C^y(r)$, which suggests an isotropic algebraic decay for long-range correlations: $C(\mathbf{r})\sim |\mathbf{r}|^{-\eta}$ for $|\mathbf{r}|\gg 1$ irrespective of its direction. Since neither the instantaneous Hamiltonian.(\ref{eq:energy}) nor the time-averaged Hamiltonian.(\ref{eq:H0}) preserve the spatial SO(2) rotational symmetry (they keep the discrete rotational symmetries instead), the SO(2) symmetry is not exact but an emergent one existing only at large scales ($|\mathbf{r}|\rightarrow\infty$). Such an enhanced symmetry is due to the fact that the non-equilibrium phase at fast driving is a critical phase, where those anisotropic perturbations are irrelevant for the physics in the long-wave limit.  Similar behavior has been discussed in the quantum and classical critical points in equilibrium systems\cite{Sreejith2019,Liu2024}.

 {\it The 1st order phase transition  v.s. phase coexistence --} The qualitatively different behavior of $C(\mathbf{r})$ in the fast and slowly driven phases indicates a phase transition between them. This exponential-to-algebraic decay and  absence of spontaneous symmetry breaking reminds us of the BKT transition in an 2D XY model. In contrast,  we find a 1st order phase transition or a phase coexistence between a metastable state and a stable state, and the  system can fall into either of them depending on its initial state.  The lifetime of the metastable state crucially depends on the system size\cite{Supplementary}, and the long-time evolution will finally drive the system to the stable state, which will
suddenly jump from the asymmetric 1D  state to the emergent isotropic 2D state via a 1st-order phase transition at a critical $T_d$.

  If the lifetime of the metastable state is much longer than the simulation time, we observe a phase coexistence with a hysteresis behavior where the value of a physical quantity  lags behind changes in the effect causing it. To verify this numerically, we slowly change $T_d$ by first increasing it  forwardly then reducing it until $T_d$ returns to its initial value. For each $T_d$ in this process, we perform simulation with the initial state as the final state obtained in the last simulation with a slightly different $T_d$ and calculate the time-averaged NN correlation $C_1$. As shown in Fig.\ref{fig:fig3} (a), a hysteresis loop appears in the $C_1-T_d$ curve where the value of $C_1$ depends on the forward or backward paths of $T_d$.

 Another important difference between this NEPT and the BKT physics of an 2D XY model is that the disordered phase in the former is actually an 1D phase, thus differs from the high-temperature phase of the 2D XY model, even though both of them are short-ranged correlated. This difference can be reflected in the temperature ($T$) dependence of the correlation lengths ($\xi$). In the slowly driven phase, the inset of Fig.\ref{fig:fig2} (b) indicates a linear relation between $1/\xi$ and T, which agrees with the 1D XY model\cite{Supplementary}, but differs from the disordered phase in the 2D XY model ($1/\xi\sim \ln T$\cite{Nagaosa1999}).

\begin{figure}[htbp]
\centering
\includegraphics[width=0.45\textwidth]{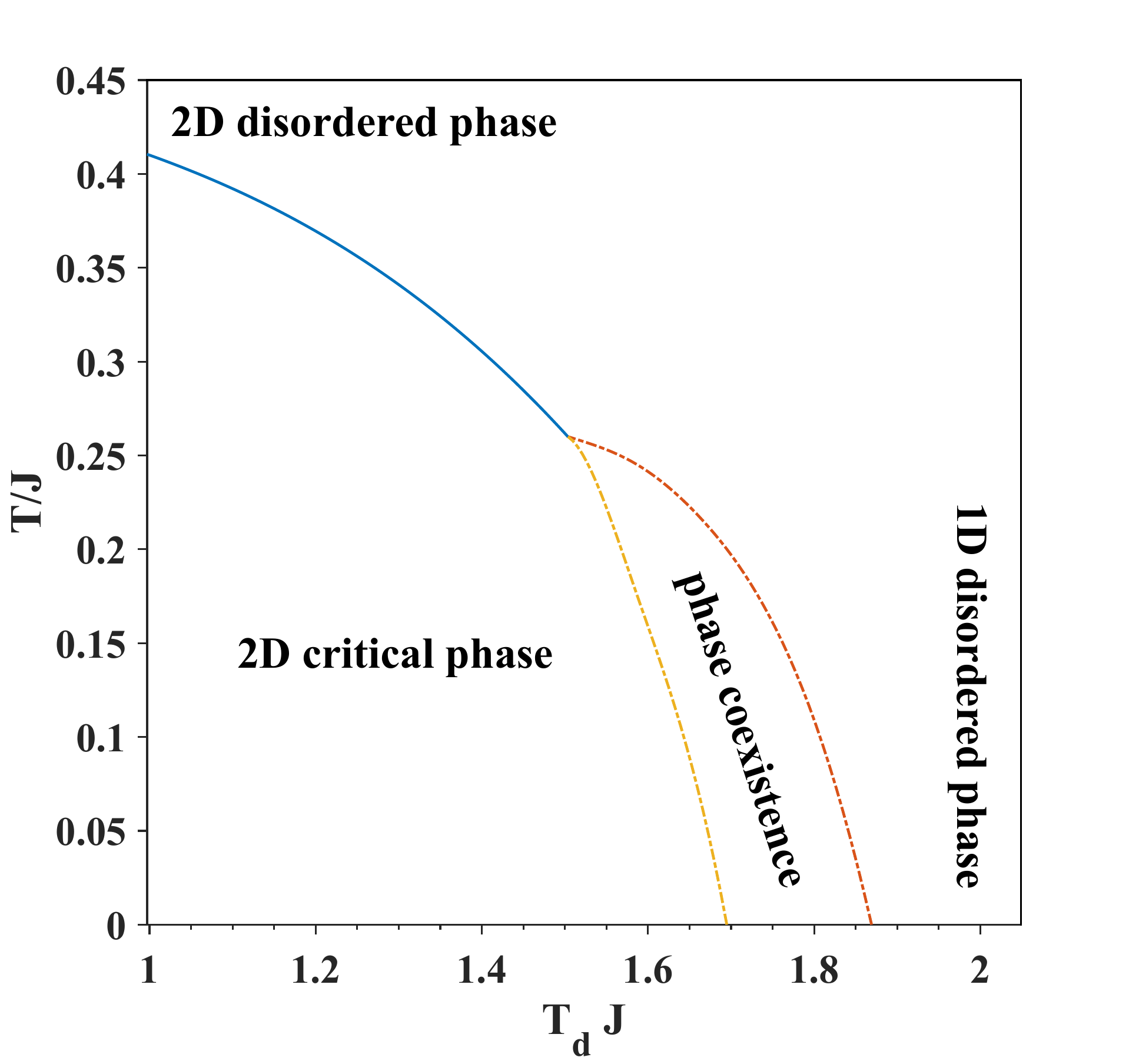}
\caption{The phase diagram of the proposed model in terms of the temperature $T$ and driving period $T_d$. $I$ and $\gamma$  are chosen the same as in Fig.\ref{fig:fig2}. }
\label{fig:fig4}
\end{figure}

{\it Finite temperature phase transitions and  phase diagram --} One may expect this emergent isotropic phase is qualitatively identical to the quasi-long-range ordered phase in a 2D XY model. However, this is not the case, at least in the intermediate driven regime, where the higher order terms in the Magus expansion could significantly influence the properties of this non-equilibrium system ({\it e.g.} its critical behavior) thus make them differ from the prediction of the zeroth-order approximation in Eq.(\ref{eq:H0}). By slowly changing the temperature along the forward-backward path, we also find a hysteresis loop in the $C_1-T$ curve as shown in Fig.\ref{fig:fig3} (b), which indicates a 1st order  instead of a BKT-type of finite-temperature phase transition. The phase diagram in terms of  $T$ and  $T_d$ is plotted in Fig.\ref{fig:fig4}, from which we can find that with faster driving, the area of the hysteresis loop shrinks and finally disappears at a tricritial point, after which the BKT phase transition is recovered\cite{Supplementary}.


{\it Experimental realizations and detections --} As for experimental realizations, the proposed model is closely related to the magnetic systems with easy-plane anisotropy. Although adjusting the lattice connectivity of a solid-state system is almot impossible, it is accessible in certain synthetic quantum materials. For example, by loading strongly interacting bosons into an optical lattice with quasi-1D cigar-like geometry, one can realized a hardcore-bosonic model whose hopping can be dynamically modulated as in Eq.(\ref{eq:Jxy}) by periodically  varying the frequency of the trap along different directions. For an 1D optical lattice with the single-particle hopping $J_h=400$Hz, one can estimate that the NEPT could be observed at a driving period $T_d\sim 10$ms, an accessible timescale in the current optical lattice setup. Another experimental proposal is based on Rydberg tweezer array.  The XY-type interaction can be realized via the coherent exchange of the internal states of the Rydberg atoms\cite{Browaeys2020}. The relative distance between the atoms can be dynamically changed by a programmable reconfiguration of the optical tweezers array, so are the interactions between the atoms. One can also estimate  the typical time scale required to observe the NEPT is a few microseconds, much shorter than the lifetime of the Rydberg atoms ($\sim 100\mu $s).

The quasi-1D and emergent 2D phases   can be observed from the time-of-flight (TOF) images of the bosons. The former will lead to anisotropic TOF images elongated along either the x or y direction, while a symmetric TOF image will be observed in the presence of fast driving.    The short-range and quasi-long-range correlations can also be distinguished from  the TOF images, where the height of coherence peak will  exhibit a  hysteresis loop around the NEPT. Although the proposed rotor model is classical, we expect that the dimension-raising NEPTs exist in this hard-core boson system regardless of its quantum nature. It is known that at sufficiently high temperature, the thermal fluctuations overwhelm the quantum fluctuations, thus the finite-temperature phase transition of a quantum model usually shares the same universality class with its classical counterpart\cite{Ceccarelli2013}. Therefore, we expect that the nature of dimension-raising NEPT will not be changed by quantum fluctuations, at least for sufficiently high temperature.

 \begin{figure}[htbp]
\centering
\includegraphics[width=0.45\textwidth]{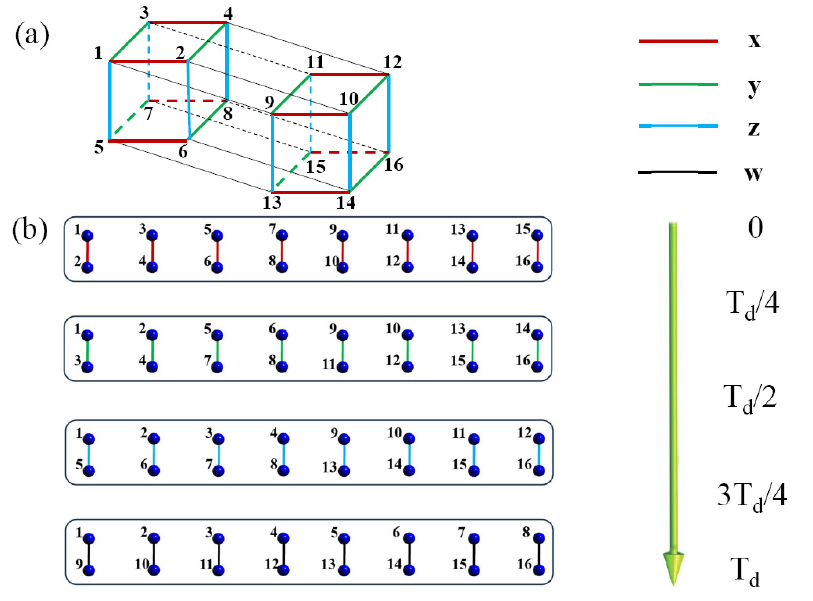}
\caption{(a) Sketches of a 4D hypercubic lattice with $N=2^4$, (b)a programable reconfiguration protocol within one driving period corresponding to Eq.(\ref{eq:Jxy2}).}
\label{fig:fig5}
\end{figure}

{\it A four-dimensional (4D) generalization --} Finally, we discuss an higher-dimensional generalization of our protocol, which enables us to realize the many-body phases with effective dimensions more than three.  A generalization to three dimension is straightforward, and its phase diagram and physical realization are similar to the 2D case\cite{Supplementary}. However, a generalization towards the effective dimensions higher than three is nontrivial, since these systems are quite rare in nature. For instance, we can  choose a time-dependent Hamiltonian:
\begin{equation}
H(t)=\sum_\mathbf{i} \big\{\frac{I}2 \dot{\theta}^2_\mathbf{i} -\sum_{\alpha=w,x,y,z} J_\alpha (t)\cos(\theta_{\mathbf{i}}-\theta_{\mathbf{i}+\hat{e}_\alpha}) \big\} \label{eq:energy4D}
\end{equation}
where $\alpha=x,y,z,w$ are the index of orthogonal directions of a 4D hypercubic lattice with total lattice sites number $L^4$ (see Fig.\ref{fig:fig5} (a) for an example with $L=2$). The time-dependent $J_\alpha(t)$ are chosen as following:


\begin{small}
\begin{equation}
\begin{cases}
J_x(t)=J,~ J_{y,z,w}(t)=0;  &  nT_d <t\leq (n+\frac 14) T_d\\
J_y(t)=J,~ J_{x,z,w}(t)=0; & ( n+\frac 14) T_d <t\leq (n+\frac 12) T_d\\
J_z(t)=J,~ J_{x,y,w}(t)=0; & ( n+\frac 12) T_d <t\leq (n+\frac 34) T_d\\
J_w(t)=J,~ J_{x,y,z}(t)=0; & ( n+\frac 34) T_d <t\leq (n+1) T_d\\
	\end{cases} \label{eq:Jxy2}
\end{equation}
\end{small}

which means that at each quarter of the driving period, the $L^4$ lattice sites are organized as $L^3$ decoupled 1D chains with length $L$ along $\alpha=x,y,z,w$ respectively (see Fig.\ref{fig:fig5} b). Such a lattice reconfiguration, even though unrealistic for the cold atom setup, is accessible in the Rydberg atom arrays, where each atom can be moved by the optical tweezers in a programable way. With fast driving, we expect the emergence of a 4D XY model. With a similar protocol, we can realize many-body phases with arbitrary effective dimensions. Furthermore, by designing more complex driving protocols, it is possible to realize the phase transitions between phases with arbitrarily different dimensions\cite{Supplementary}.

{\it Conclusion and outlook --} As a conclusion, we propose a mechanism which allows us to dynamically adjust the effective dimension of many-body systems. It differs from intensively-studied dimensional crossover phenomena with tunable anisotropic inter-chain or inter-layer couplings\cite{Klanjsek2008,Vogler2014,Revelle2016,Moller2021,Yao2023,Shah2023,Kirankumar2024}, in the sense that the high-dimensional physics in our case emerges spontaneously via a phase transition instead of crossover. Due to its nonequilibrium feature, the proposed mechanism is also different from those governing the dimensional reduction phenomena, which are induced either by the presence of anisotropic orbit order\cite{Khomskii2005,Streltsov2017}, or by the geometric frustration\cite{Sebastian2006,Batista2007,Okuma2021}.

Since the dimensionality is an emergent property in our driving protocols, one may wonder whether other important concepts, such as the frustration, can emerge in a similar way. If the lattice connectivity periodically switches between two frustration-free configurations, the system adiabatically follows their respective equilibrium states for a sufficiently slow driving. However, as the driving frequency exceeds a critical value, frustration  spontaneously emerges if the average Hamiltonian between those two frustration-free configurations are frustrated. It would be interesting to investigate the properties of such  emergent frustrated systems ({\it e.g.} the order-by-disorder phenomenon, fractionalized excitations), and compared them to their equilibrium counterparts.

{\it Acknowledgments}.--- This work is supported by the National Key Research and Development Program of China (2020YFA0309000,2024YFA1408303), Natural Science Foundation of China (Grant No.12174251) and Shanghai (Grant No.22ZR142830),  Shanghai Municipal Science and Technology Major Project (Grant No.2019SHZDZX01), Shanghai Science and Technology Innovation Action Plan(Grant No. 24Z510205936). ZZC is also supported by the National Center for High-Level Talent Training in Mathematics, Physics, Chemistry, and Biology.


\end{document}